\documentclass[a4paper,11pt]{article}
\usepackage{pos}

\usepackage{xcolor}

\usepackage{braket}
\usepackage{graphicx}
\usepackage{amsmath}

\title{First Lattice QCD Study of Proton Twist-3 GPDs}

\author[a]{S. Bhattacharya}
\author[b]{K. Cichy}
\author[a]{M. Constantinou}
\author*[a]{J. Dodson}
\author[a]{A. Metz}
\author[a]{A. Scapellato}
\author[c]{F. Steffens}

\affiliation[a]{Department of Physics,  Temple University,  \\ Philadelphia,  PA 19122 - 1801,  USA}
\affiliation[b]{Faculty of Physics, Adam Mickiewicz University, \\ ul.\ Uniwersytetu Pozna\'nskiego 2, 61-614 Pozna\'{n}, Poland}
\affiliation[c]{Institut f\"ur Strahlen- und Kernphysik, Rheinische
  Friedrich-Wilhelms-Universit\"at Bonn, \\ Nussallee 14-16, 53115 Bonn}

\emailAdd{tug76350@temple.edu}

\abstract{We present first results on selected twist-3 quark GPDs using the quasi-distribution method. This approach relates lattice QCD data and light-cone distribution functions using Large Momentum Effective Theory (LaMET). 
We calculate quark-antiquark correlators of boosted nucleons coupled to non-local operators with vector and axial Dirac structure, which is transverse to the momentum boost. We use three values of the momentum boost, namely 0.83, 1.25, 1.67 GeV. The GPDs are defined in the symmetric (Breit) frame, which we implement here with 4-vector momentum transfer squared of 0, 0.69 and 1.39 GeV$^2$, all at zero skewness. The calculation is performed using one ensemble of two degenerate light, a strange and a charm quark ($N_f=2+1+1$) of maximally twisted mass fermions with a clover term, corresponding to a pion mass of 260 MeV.
}

\FullConference{%
 The 38th International Symposium on Lattice Field Theory, LATTICE2021
  26th-30th July, 2021
  Zoom/Gather@Massachusetts Institute of Technology
}

\begin{document}
\maketitle

\section{Introduction}

One of the key determinants in proton structure are distribution functions of its partonic content, which are extensively studied experimentally worldwide in major laboratories, such as JLab, BNL, Fermilab, DESY, SLAC, CERN, PSI, J-PARC, and MAMI. The key quantities for mapping the proton structure are  parton distribution functions (PDFs), generalized parton distributions (GPDs), and transverse-momentum-dependent distributions (TMDs). These are inferred from experimental data from high-energy scattering processes, which is possible due to the asymptotic freedom of the strong coupling. In particular, asymptotic freedom enables the use of a QCD factorization formalism to isolate the universal non-perturbative component of the process, namely the PDFs, GPDs, and TMDs. Furthermore, these distribution functions are classified based on their twist, which indicates the order at which they appear in the expansion in terms of the large energy, $Q$, of the physical process. At leading twist (twist-2), they have probabilistic interpretation, which does not hold for the higher-twist parton distributions. The twist-2 contributions have been extensively studied, while very little is known about the twist-3 contributions. However, twist-3 contributions are not negligible for the energy scales explored experimentally; in fact, they are assumed to be sizable, which we address in this work. Twist-3 GPDs are also interesting in their own right. For instance, the twist-3 GPDs that we present here are necessary for proton tomography and for exploring the spin-orbit correlations in the proton~\cite{Lorce}. This means that one must know the twist-3 GPDs in order to reliably extract the twist-2 contributions. Knowledge of the twist-3 GPDs can be used to estimate the power corrections in hard exclusive processes, such as DVCS. Despite their importance, higher-twist distributions can be difficult to determine experimentally, since it is challenging to isolate them from the leading-twist contributions. All the above are important motivations to undertake calculations of twist-3 GPDs from lattice QCD, which, while very challenging, are very promising in providing information on these quantities.

\section{$x$-dependent GPDs from lattice QCD}

The light-cone nature of PDFs, GPDs, and TMDs makes them directly inaccessible from lattice QCD that is based on a Euclidean metric. However, the development of approaches other than Mellin moments has led to a successful research program to obtain the $x$-dependence of distribution functions from lattice QCD. A widely used approach is based on calculations of matrix elements of non-local operators and boosted protons, the so-called quasi-distributions method. Such matrix elements are then related to the light-cone distributions via Large-Momentum Effective Theory (LaMET)~\cite{Ji:2013dva}. Extensive reviews of the various approaches to get $x$-dependent distribution functions can be found in Refs.~\cite{Cichy:2018mum,Ji:2020ect,Constantinou:2020pek,Cichy:2021lih}. 

Here, we implement the quasi-distribution method to obtain information on twist-3 GPDs. There are several computational challenges associated with this calculation, mainly due to the use of matrix elements of boosted proton states and non-local operators at off-forward kinematics. First, due to the momentum transfer, there are increased statistical uncertainties compared to the PDFs case. Second, since the GPDs are defined in the Breit frame, each value of momentum transfer requires a separate calculation. Third, as we shall see, there is a need for as many independent matrix elements as there are GPDs, so that we can disentangle them. Last but not least, the matching kernel is more complicated for the case of nonzero skewness. In fact, the perturbative formalism breaks down around $x = |\xi|$ due to severe higher-twist contributions. Even at $\xi=0$, a proper matching formalism for the twist-3 GPDs is much more complicated than for twist-2 GPDs due to the presence of zero modes, as well as mixing with quark-gluon-quark correlators. In this work, we neglect such a mixing. Furthermore, the matching kernel for the twist-3 vector GPDs is not known, and in this preliminary work we only present results for the vector matrix elements. For the twist-3 axial GPDs, we use the matching obtained for twist-3 PDFs. Similar to the case of twist-2 GPDs, it is anticipated that the matching of twist-3 GPDs at $\xi=0$ is the same as for twist-3 PDFs.
For the twist-3 GPDs under study, we construct the matrix elements
\begin{equation}
\label{eq:ME}
 h_{{\cal O}} ( \Gamma_\kappa, z, P_f, P_i, \mu) = Z_{\cal O}(z,\mu) \bra{N(P_f)} \overline{\psi}(z) \, {\cal O} \,\mathcal{W}(z,0) \psi(0) \ket{N (f)}\,,
\end{equation}
where, as in the twist-2 case, the Wilson line is in the same direction as the boost, which is chosen to be the $z$-direction. $\Gamma_\kappa$ is the parity projector with $\kappa=0$ representing the unpolarized and $\kappa=1,\,2,\,3$ the polarized projector. The operator ${\cal O} $ corresponding to the twist-3 contributions are $\gamma^j$ and $\gamma^5\,\gamma^j$, both with $j=1,2$. GPDs require $\mathbf{P_f} - \mathbf{P_i} = \mathbf{\Delta} \ne 0$, and are defined in the Breit frame, in which $\mathbf{P}_f = \mathbf{P} + \frac{\mathbf{\Delta}}{2}$ and $\mathbf{P}_i= \mathbf{P} - \frac{\mathbf{\Delta}}{2}$, where $\mathbf{P} = (0,0,P_3)$ is the proton momentum. Note that GPDs depend on the 4-vector momentum transfer squared, $t$, and not on the individual nucleon momenta, while the matrix elements depend on the source and sink momenta. Another important variable of GPDs is the skewness $\xi$, which is related to the momentum transfer in the direction of the boost. On the lattice, we define the quasi-skewness, $\xi=-\frac{\Delta_3}{2P_3}$, and in our calculation we focus on $\xi=0$.

The matrix elements of Eq.~\eqref{eq:ME} are renormalized non-perturbatively in the RI' scheme and are converted to the modified $\overline{\rm MS}$ (M$\overline{\rm MS}$) scheme and evolved to a scale of 2 GeV.
These renormalized matrix elements decompose into four GPDs for the vector ($G_1$, $G_2$, $G_3$, $G_4$) and four for the axial ($\widetilde{G}_1$, $\widetilde{G}_2$, $\widetilde{G}_3$, $\widetilde{G}_4$) case. Therefore, one needs four independent matrix elements to disentangle them. These are obtained from all non-vanishing combinations of the Dirac indices ($j=1,2$) and parity projectors ($\kappa=0,1,2,3)$. This decomposition is applied at each value of $z$ separately, and the relevant expressions are~\cite{Kiptily:2002nx, Aslan:2018zzk}.
\begin{eqnarray}
\label{eq:v_decomp}
\hspace*{-0.4cm}
h_{\gamma^j} \hspace*{-0.25cm}&=& \hspace*{-0.25cm}
\langle\langle \frac{g_\perp^{j\rho} \Delta_\rho}{2m} \rangle\rangle [F_E {+} F_{G_1}] + 
\langle\langle g_\perp^{j\rho} \gamma_\rho \rangle\rangle [F_H {+} F_{G_2}] + 
\langle\langle \frac{g_\perp^{j\rho} \Delta_\rho \gamma^+}{P^+} \rangle\rangle F_{G_3} + 
\langle\langle \frac{i \epsilon_\perp^{j\rho} \Delta_\rho \gamma^+ \gamma_5}{P^+} \rangle\rangle F_{G_4} \,, \hspace*{0.5cm} \\[1ex]
 \label{eq:a_decomp}
 \hspace*{-0.4cm}
 h_{\gamma^j\gamma_5} \hspace*{-0.25cm}&=& \hspace*{-0.25cm}
\langle\langle \frac{g_\perp^{j\rho} \Delta_\rho \gamma_5}{2m} \rangle\rangle [F_{\widetilde{E}} + F_{\widetilde{G}_1}] + 
\langle\langle  g_\perp^{j\rho} \gamma_\rho \gamma_5\rangle\rangle [F_{\widetilde{H}} + F_{\widetilde{G}_2}] + 
\langle\langle \frac{g_\perp^{j\rho} \Delta_\rho \gamma^+ \gamma_5}{P^+} \rangle\rangle F_{\widetilde{G}_3} + 
\langle\langle \frac{i \epsilon_\perp^{j\rho} \Delta_\rho  \gamma^+}{P^+} \rangle\rangle F_{\widetilde{G}_4} \,,
\hspace*{0.5cm}
\end{eqnarray}
where $\langle\langle \Gamma \rangle\rangle \equiv \bar{u}_N(P_f,s')\, \Gamma \,u_N(P_i,s)$ with $u_N$ the proton spinors. For simplicity, we omit the arguments of the matrix elements, that is, $ h_{{\cal O}} \equiv h_{\cal O} ( \Gamma_\kappa, z, P_f, P_i, \mu)$, and $ F_X \equiv F_X(z,  \xi, t, P_3,\mu)$.  $F_{{H}},\,F_{{E}}$, $F_{\widetilde{H}}$, and $F_{\widetilde{E}}$ are twist-2 contributions, while $F_{\widetilde{G}_i}$ are the twist-3 contributions. Based on the above decomposition, the forward limit of $[F_{\widetilde{H}} + F_{\widetilde{G}_2}]$ is the twist-3 PDF $g_T$. It should be noted that the forward limit of the vector case is zero.
 
The functions $F_X$ are defined in coordinate space and one must Fourier-transform them to momentum space,
 \begin{equation}
 G_{X,{\rm Q}}(x,\xi,t,P_3,\mu) = \int\frac{dz}{4\pi} e^{-ixP_3z}F_X(z,  \xi, t, P_3,\mu)\,.
   \end{equation} 
With the values of $z$ being limited up to half of the spatial extent of the lattice, the reconstruction of the $x$-dependence is an ill-posed inverse problem that does not have a unique solution. To avoid imposing assumptions like $F_X$ being zero beyond some $z_{\rm max}$, we use the model-independent Backus-Gilbert reconstruction method. More details on the implementation can be found in~\cite{Alexandrou:2020zbe, alexandrou2021transversity}. The last step of the calculation is to match the quasi-GPDs, $G_{X,{\rm Q}}$, to the light-cone GPDs, $G_X$, via a kernel calculated in perturbation theory,
  \begin{equation}
     G_{X,{\rm Q}}(x,\xi,t,P_3,\mu) = \int_{-1}^1 \frac{dy}{|y|} C_X\left(\frac{x}{y},\frac{\xi}{y},\frac{\mu}{yP_3},r\right) G_X(y,\xi,t,\mu) + \mathcal{O}\left(\frac{m^2}{P_3^2}, \frac{t}{P_3^2}, \frac{\Lambda_QCD^2}{x^2P_{3_{\phantom{L}}}^2}\right)\,.
      \end{equation}
For the matching kernel, $C_X$, we use the one-loop expression from Ref.~\cite{Bhattacharya:2020xlt} for the axial GPDs. The formula we use connects the quasi-distributions in the M$\overline{\rm MS}$ scheme to the light-cone distributions in the standard $\overline{\rm MS}$ scheme. The renormalization scale is chosen to be 2 GeV for both quantities. The corresponding kernel for the vector case is not known yet, and therefore, we do not have results on the light-cone GPDs $G_1$, $G_2$, $G_3$, and $G_4$.

\section{Lattice calculation} 
 
The workflow of this calculation follows the procedure used for the unpolarized, helicity and transversity GPDs~\cite{PhysRevLett.125.262001,alexandrou2021transversity}. We start by constructing the two-point and three-point functions for the nucleon,
\begin{equation}
 C^{2pt}(\textbf{P}, t_s, 0) = \Gamma_{0\alpha\beta} \sum_\textbf{x} e^{-i \textbf{P} \cdot \textbf{x}} \bra 0 N_\alpha(\textbf{x}, t_s) N_\beta(\textbf{0},0)\ket 0\,,  \\
\end{equation}
\begin{equation}
 C^{3pt}_{\Gamma_\nu} (\mathbf{P_f}, \mathbf{P_i}, t_s, \tau, 0)  = \Gamma_{\nu \alpha \beta} \sum_{\mathbf{x},\mathbf{y}} e^{-i (\mathbf{P_f} - \mathbf{P_i}) \cdot \mathbf{y}} e^{-i \mathbf{P_f} \cdot \mathbf{x}} \bra 0 N_\alpha(\mathbf{x}, t_s)\mathcal{O}_{\gamma^i \gamma^5}(\mathbf{y},\tau;z) N_\beta(\mathbf{0},0)\ket 0\,.
\end{equation}
Here, $N_\alpha,N_\beta$ are the interpolating fields for the proton, $\tau$ is the current insertion time, and $t_s$ is the time separation between the source and the sink (the source is taken at $t=0$). $\Gamma_0 = \frac{1 + \gamma_4}{2}$ is the parity plus projector and $\Gamma_\kappa$ is either $\Gamma_0$ or $\Gamma_j = \frac{1}{4}(1 + \gamma^0)i\gamma^5\gamma^j$. 

\begin{figure}[h!]
\begin{minipage}{5cm} 
\includegraphics[scale=0.85]{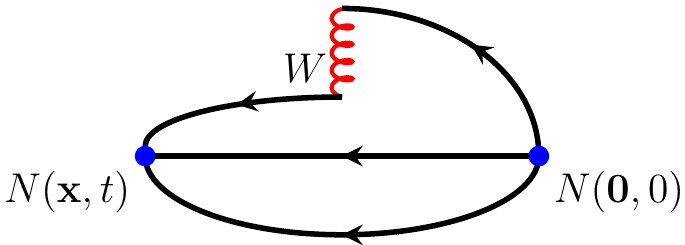}
\end{minipage}
\hfill
\begin{minipage}{8.5cm}
\caption{Pictorial representation of the connected contributions to the three-point functions. The initial and final states with the quantum numbers of the nucleon are indicated by $N(\mathbf{0},0)$ and $N(\mathbf{x},t)$, respectively. The red curly line indicates the Wilson line, $W$, of the non-local operator.}
\label{fig:diagram}
\end{minipage}
\end{figure}

We calculate connected contributions to the three-point functions, as we focus on the $u-d$ isovector flavor combination, for which the disconnected component is zero. The connected diagram is shown in Fig.~\ref{fig:diagram}.

Since there is a non-zero momentum transfer between the initial and final states, one has to cancel the time dependence of the exponentials as well as the overlap from the interpolating fields. To this end, we build the following ratio
\begin{equation}
R_{\cal O} (\Gamma_\kappa, P_f, P_i; t_s, \tau) = \frac{C^{\rm 3pt}_{\cal O} (\Gamma_\kappa, P_f, P_i; t_s, \tau)}{C^{\rm 2pt}(\Gamma_0, P_f;t_s)} \sqrt{\frac{C^{\rm 2pt}(\Gamma_0, P_i, t_s-\tau)C^{\rm 2pt}(\Gamma_0, P_f, \tau)C^{\rm 2pt}(\Gamma_0, P_f, t_s)}{C^{\rm 2pt}(\Gamma_0, P_f, t_s-\tau)C^{\rm 2pt}(\Gamma_0, P_i, \tau)C^{\rm 2pt}(\Gamma_0, P_i, t_s)}}\,.
 \end{equation}
Ground-state dominance is achieved as $t_s$ increases, with the ratio $R_{\cal O}$ becoming constant.  We extract the ground-state matrix elements by taking a plateau fit in  a region of convergence, which we indicate by $\Pi({\cal O},\Gamma_\kappa)$. For simplicity, the dependence on $z$, $P_f,\, P_i$ and the renormalization scale $\mu$ is implied.

Our calculations are performed on an $N_f=2+1+1$ ensemble~\cite{Alexandrou:2021gqw} of maximally twisted mass fermions, with pion mass $M_\pi=260$~MeV, lattice spacing $a\simeq 0.093$~fm and volume $V~=~32^3\times~64$. 
We produce data for $h_{\gamma^1}$, $h_{\gamma^2}$, $h_{\gamma^1\gamma_5}$ and $h_{\gamma^2\gamma_5}$, for a class of momenta of the form $\mathbf{\Delta} = (\pm q,0,0)$, $\mathbf{\Delta} = (0,\pm q,0)$, and $\mathbf{\Delta} = (\pm q,\pm q,0)$. The nucleon boost is nonzero along the $z$-direction, $\mathbf{P}~=~(0,0,\pm P_3)$. This leads to a factor of eight more statistics. The parameters of the calculation and the number of measurements is given in Table~\ref{tab:params}. The source-sink time separation is chosen as $t_s=10 a$, due to the increased statistical uncertainties in the twist-3 contributions compared to the twist-2 case.

\begin{table}[h!]
\centering
\begin{tabular}{|c|c|c|c|c|c|}
\hline
$P_3\, [\,{\rm GeV}\,]$ & $q\, [\,\frac{2\pi}{L}\,]$  & $-t\, [\,{\rm GeV}^2\,]$ & $N_{\rm confs}$ & $N_{\rm src}$ & $N_{\rm total}$ \\ \hline
$\pm 0.83$ & $(\pm 2,0,0)$      & 0.69  &67   &8  & 4288 \\ \hline
$\pm 1.25$ & $(\pm 2,0,0)$      & 0.69  &67   &8  & 4288 \\ \hline
$\pm 1.25$ & $(\pm 2,\pm 2,0)$  & 1.39  &198  &8  & 12672 \\ \hline
$\pm 1.67$ & $(\pm 2,0,0)$      & 0.69  &219  &32 & 56064 \\ \hline
\end{tabular}
\caption{Statistics used in this calculation ($\xi=0$). $N_{\rm confs}$, $N_{\rm src}$ and $N_{\rm total}$ is the number of configurations, source positions per configuration and total statistics (including a factor of 8), respectively.}
    \label{tab:params}
\end{table}
 
\section{Results}

Let us begin our discussion with the bare matrix elements for the ground state, as extracted from the plateau fit. 
First, we give an example of the independent matrix elements that can be used to disentangle the GPDs, based on the trace algebra of Eqs.~\eqref{eq:v_decomp} - \eqref{eq:a_decomp}. For the kinematic setup $Q=(q,0,0)$, each one of the vector matrix elements contributes to the following GPDs: 
$\Pi (\gamma^1 , \Gamma_0)$: $E+G_1$ and  $G_3$; 
$\Pi (\gamma^1 , \Gamma_2)$: $E+G_1$ and  $G_3$; 
$\Pi (\gamma^2 , \Gamma_3)$: ${H}+{G}_2$ and ${G}_4$. As can be seen, the standard unpolarized and polarized projectors are not sufficient to isolate ${H}+{G}_2$. We are now exploring an alternative setup.
For the axial case, we have the following contributions.
$\Pi (\gamma^2 \gamma^5, \Gamma_0)$: $\widetilde{H}+\widetilde{G}_2$ and  $\widetilde{G}_4$; 
$\Pi (\gamma^2 \gamma^5, \Gamma_2)$: $\widetilde{H}+\widetilde{G}_2$ and $\widetilde{G}_4$;
$\Pi (\gamma^1 \gamma^5, \Gamma_1)$: $\widetilde{H}+\widetilde{G}_2$ and $\widetilde{E} + \widetilde{G}_1$;
$\Pi (\gamma^1 \gamma^5, \Gamma_3)$: $\widetilde{G}_3$.

\begin{figure}[h!]
\includegraphics[scale=0.5]{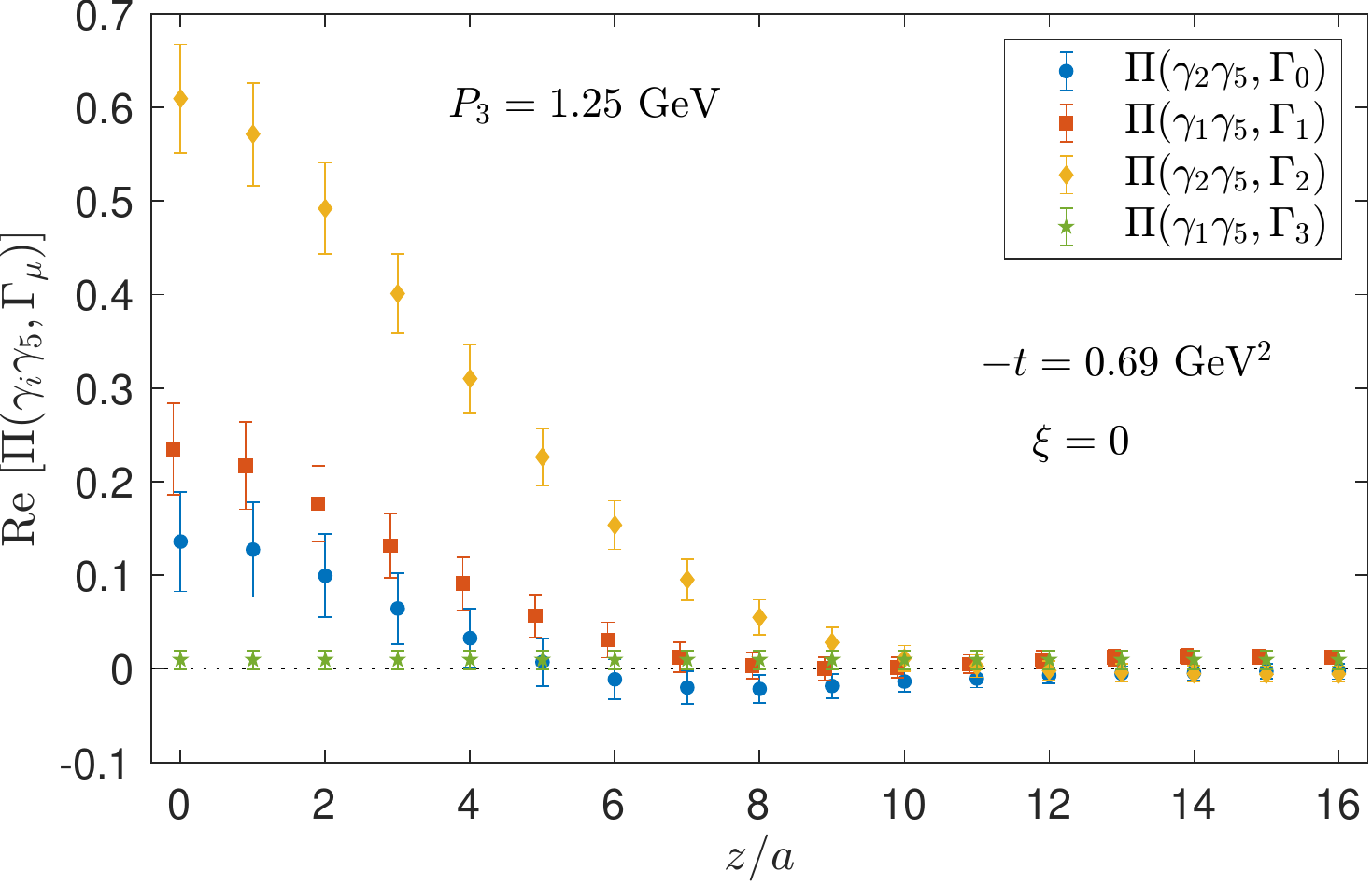}
\includegraphics[scale=0.5]{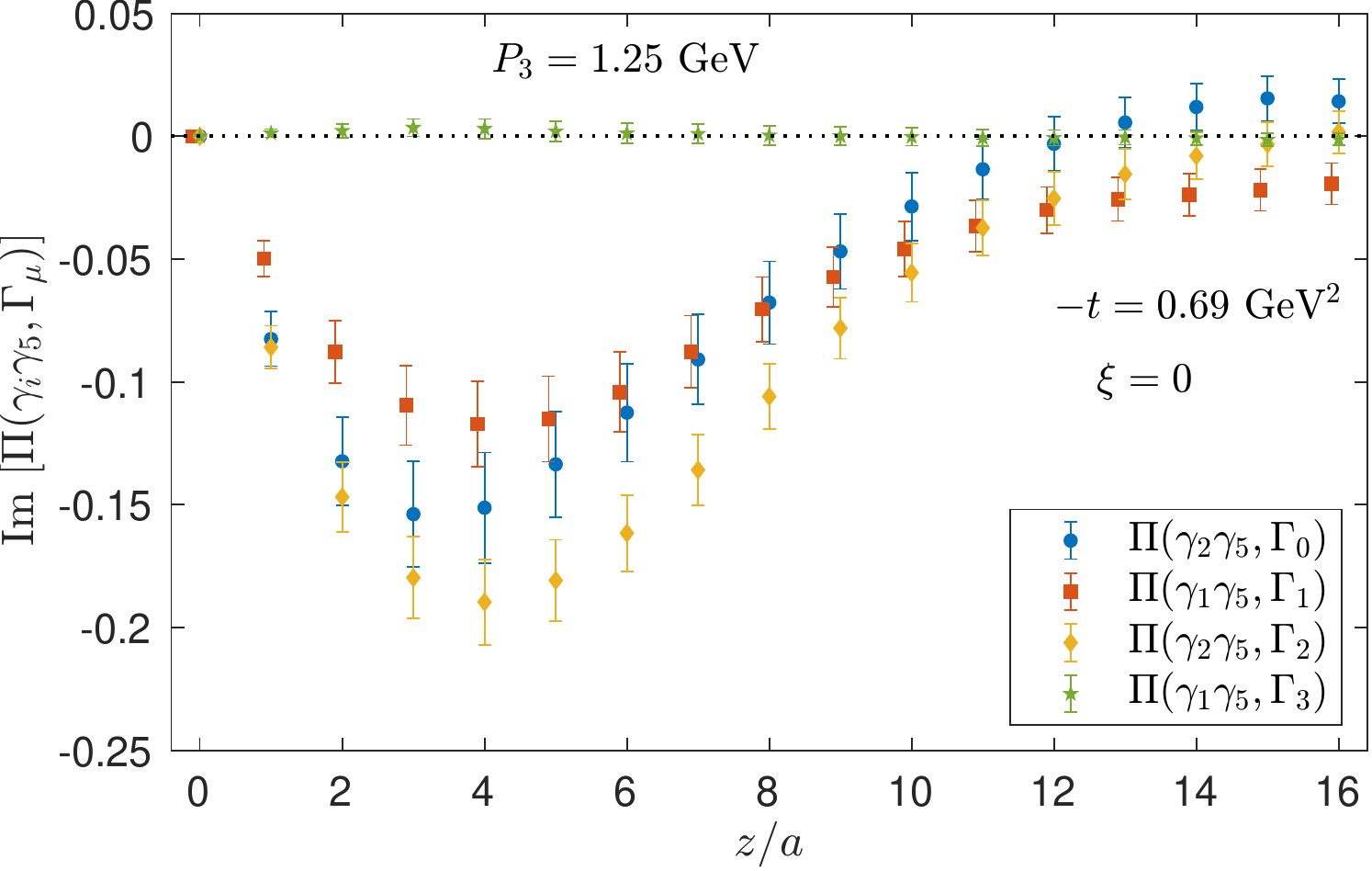}
\caption{Matrix elements contributing to the twist-3 axial GPDs for $P_3=1.25$ GeV, $t=-0.69$ GeV$^2$ and $\xi=0$. The blue, red, orange, and green points correspond to $\Pi (\gamma^2 \gamma^5, \Gamma_0)$, $\Pi (\gamma^1 \gamma^5, \Gamma_1)$, $\Pi (\gamma^2 \gamma^5, \Gamma_2)$, $\Pi (\gamma^1 \gamma^5, \Gamma_3)$.}
\label{fig:ME_A}     
\end{figure}

\begin{figure}[h!]
\includegraphics[scale=0.5]{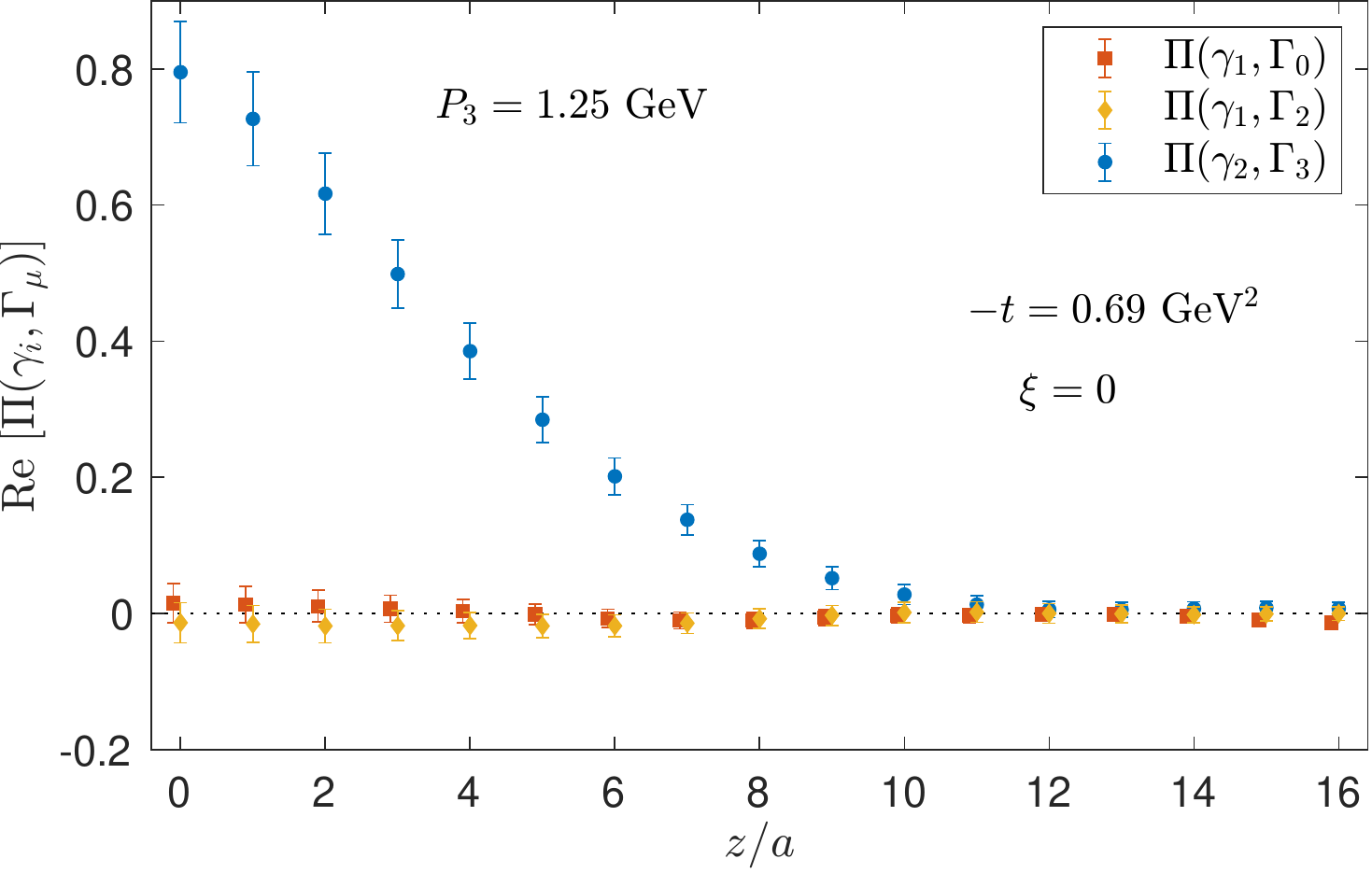}
\includegraphics[scale=0.5]{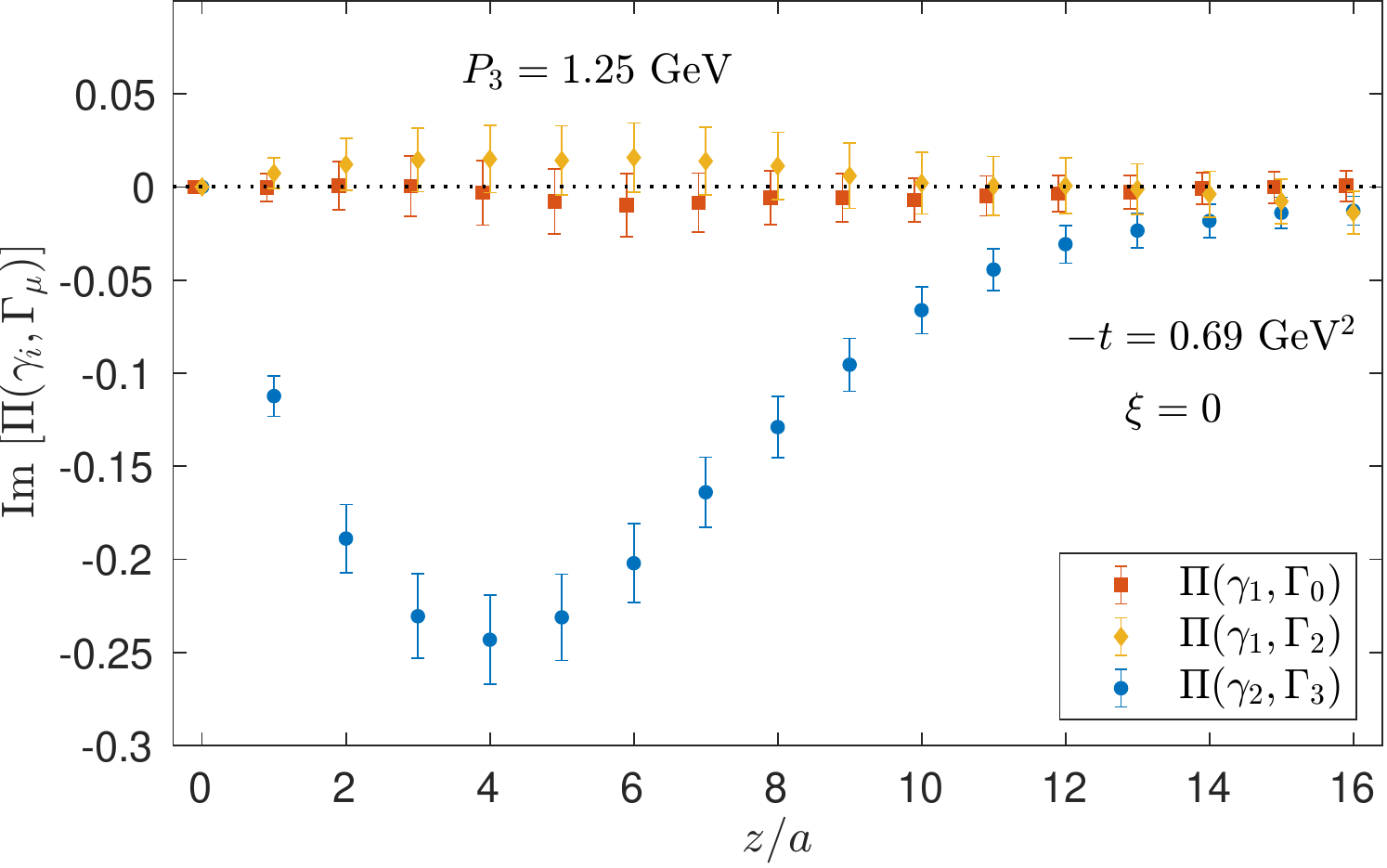}
\caption{Matrix elements contributing to the twist-3 vector GPDs for $P_3=1.25$ GeV, $t=-0.69$ GeV$^2$ and $\xi=0$. The red, orange and blue points correspond to $\Pi (\gamma^1, \Gamma_0)$, $\Pi (\gamma^1, \Gamma_2)$, $\Pi (\gamma^2, \Gamma_3)$.}
\label{fig:ME_V}     
\end{figure}

In Fig.~\ref{fig:ME_A}, we plot the four independent matrix elements contributing to the axial twist-3 GPDs for $t = -0.69$ GeV$^2$ and $P_3 = 1.25$ GeV. We have already averaged the eight combinations that lead to the same equation. When combining such matrix elements, we apply a weighted average. We use this momentum for demonstration purposes, because the small uncertainties allow one to comment on the differences in the matrix elements.
We find a very good signal for the axial case, and observe that $\Pi(\gamma^2 \gamma^5 ,\Gamma_2)$ is dominant in magnitude. $\Pi(\gamma^1 \gamma^5 ,\Gamma_3)$ is suppressed and compatible with zero. This remains true for higher values of the momentum boost. This impacts directly the extraction of $\widetilde{G}_3$, as the particular matrix element is proportional to $F_{\widetilde{G}_3}$. The matrix elements $\Pi (\gamma^1 \gamma^5, \Gamma_1)$ and $\Pi (\gamma^2 \gamma^5,\Gamma_0)$ are of similar magnitude. Results on the vector matrix elements are shown in Fig.~\ref{fig:ME_V} for the three matrix elements that contribute for $(\pm q,0,0)$ and $(0,\pm q,0)$, after averaging over all eight combinations. We find that $\Pi (\gamma^2, \Gamma_3)$ gives a sizeable signal, while the other two matrix elements are compatible with zero. However, for the vector case, further investigation is required before being able to disentangle the vector GPDs.

\begin{figure}[h!]
\includegraphics[scale=0.51]{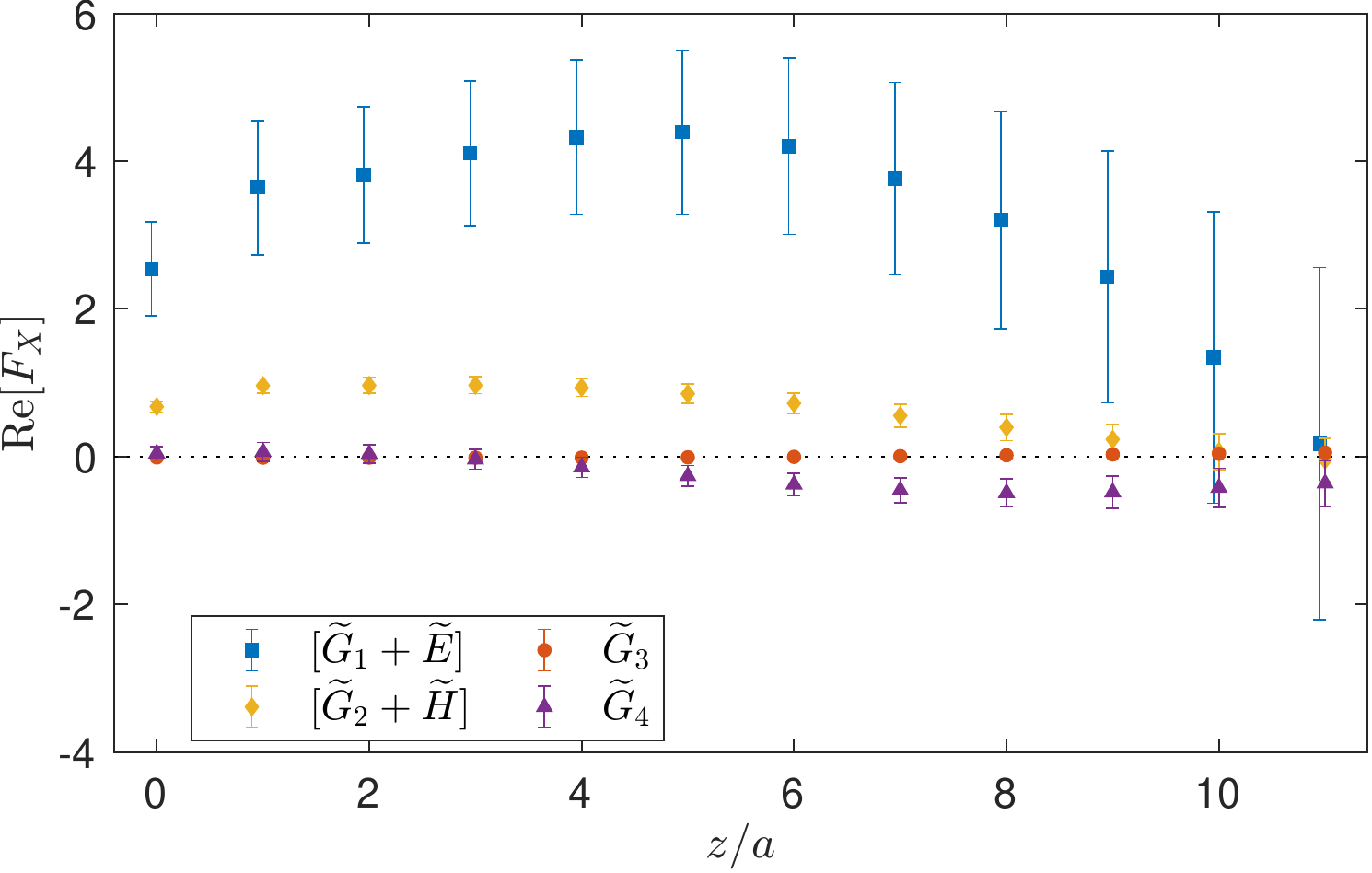}\hspace{0.1cm}
\includegraphics[scale=0.51]{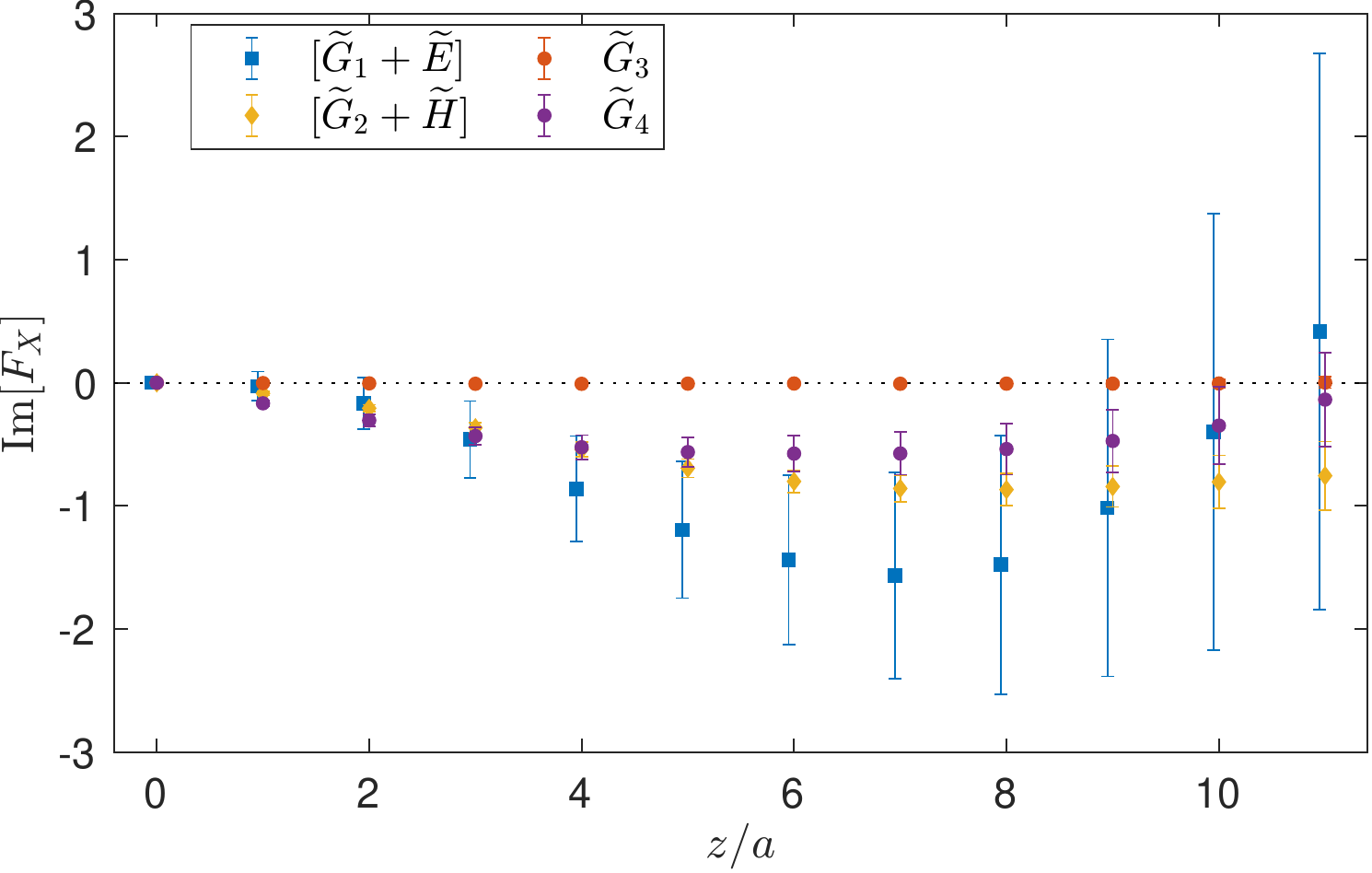}
\caption{Decomposed $F_X$ for the axial GPDs, at $\lbrace \xi=0,-t=0.69$ GeV$^2\rbrace$. The nucleon boost is $P_3~=~1.25$~GeV. The blue, orange, red, and purple points correspond to $\widetilde{H}+\widetilde{G}_2,\,\widetilde{E}+\widetilde{G}_1,\,\widetilde{G}_3,\,\widetilde{G}_4$, respectively.}
\label{fig:FX}
\end{figure}

It is also interesting to see the decomposed functions $F_X$ for $X=\widetilde{H}+\widetilde{G}_2,\,\widetilde{E}+\widetilde{G}_1,\,\widetilde{G}_3,\,\widetilde{G}_4$, in particular their dependence on the momentum boost and momentum transfer. Representative results are shown in Fig.~\ref{fig:FX}. 
As expected, the contribution from $\widetilde{E}+\widetilde{G}_1$ has the largest magnitude, followed by $\widetilde{H}+\widetilde{G}_2$. This is in accordance with the findings of the twist-2 case~\cite{Alexandrou:2020zbe}, as well as of the usual axial form factors~\cite{Alexandrou:2020okk}. The values for $\widetilde{G}_3$ are found to be exactly zero, and $\widetilde{G}_4$ has a zero real part for $z\le6a$. We note that the integrals $\int dx \widetilde{G}_i=0$ ($i=1,2,3,4$)~\cite{Kiptily:2002nx}, which might explain why $\widetilde{G}_4$ which appears on its own in the decomposition, is very small. In addition, $\int dx x \widetilde{G}_3=\frac{\xi}{4} G_E(t)$, which is zero in our calculation ($\xi=0$). 

\begin{figure}[h!]
\includegraphics[scale=0.52]{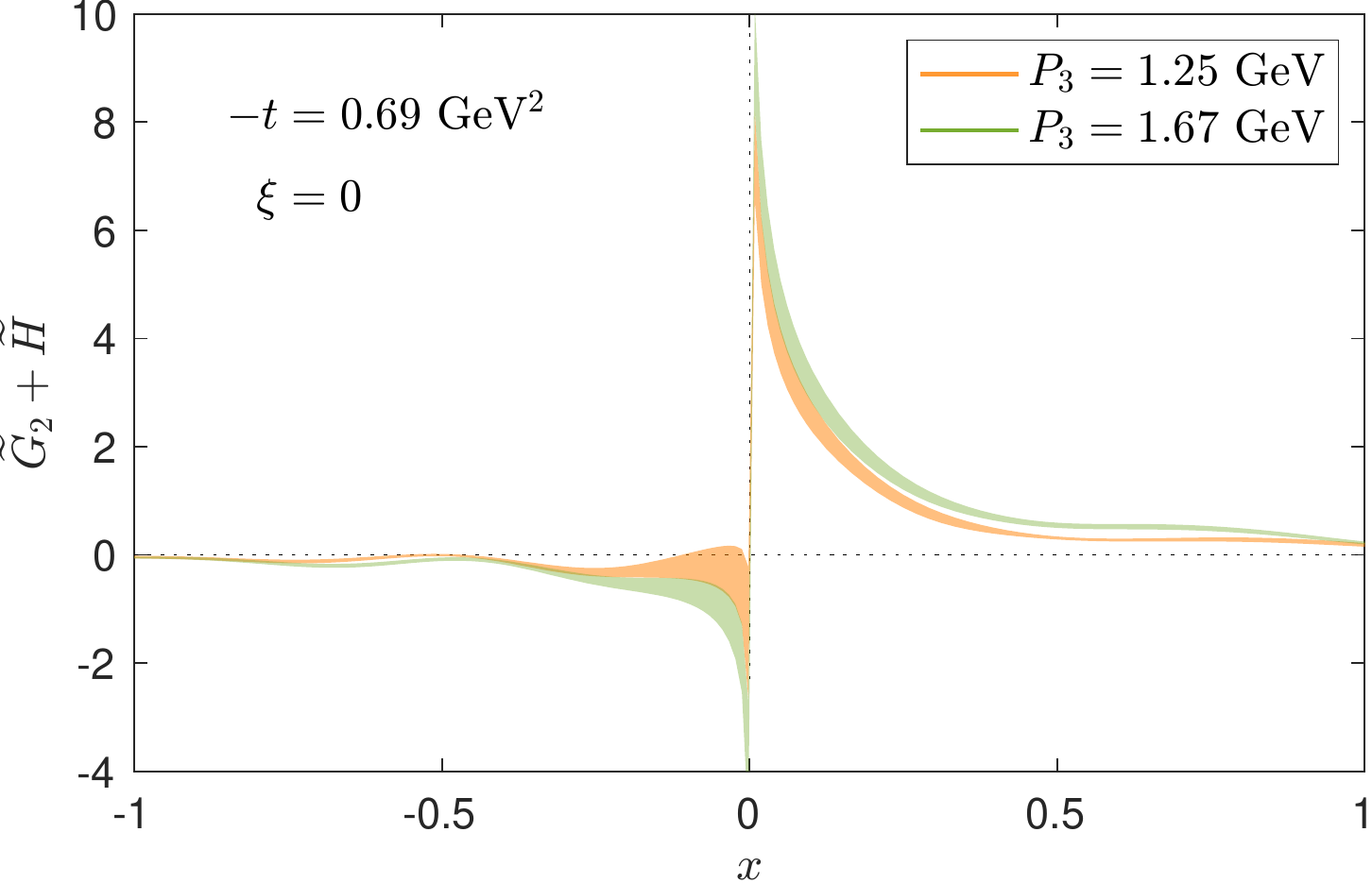}
\includegraphics[scale=0.52]{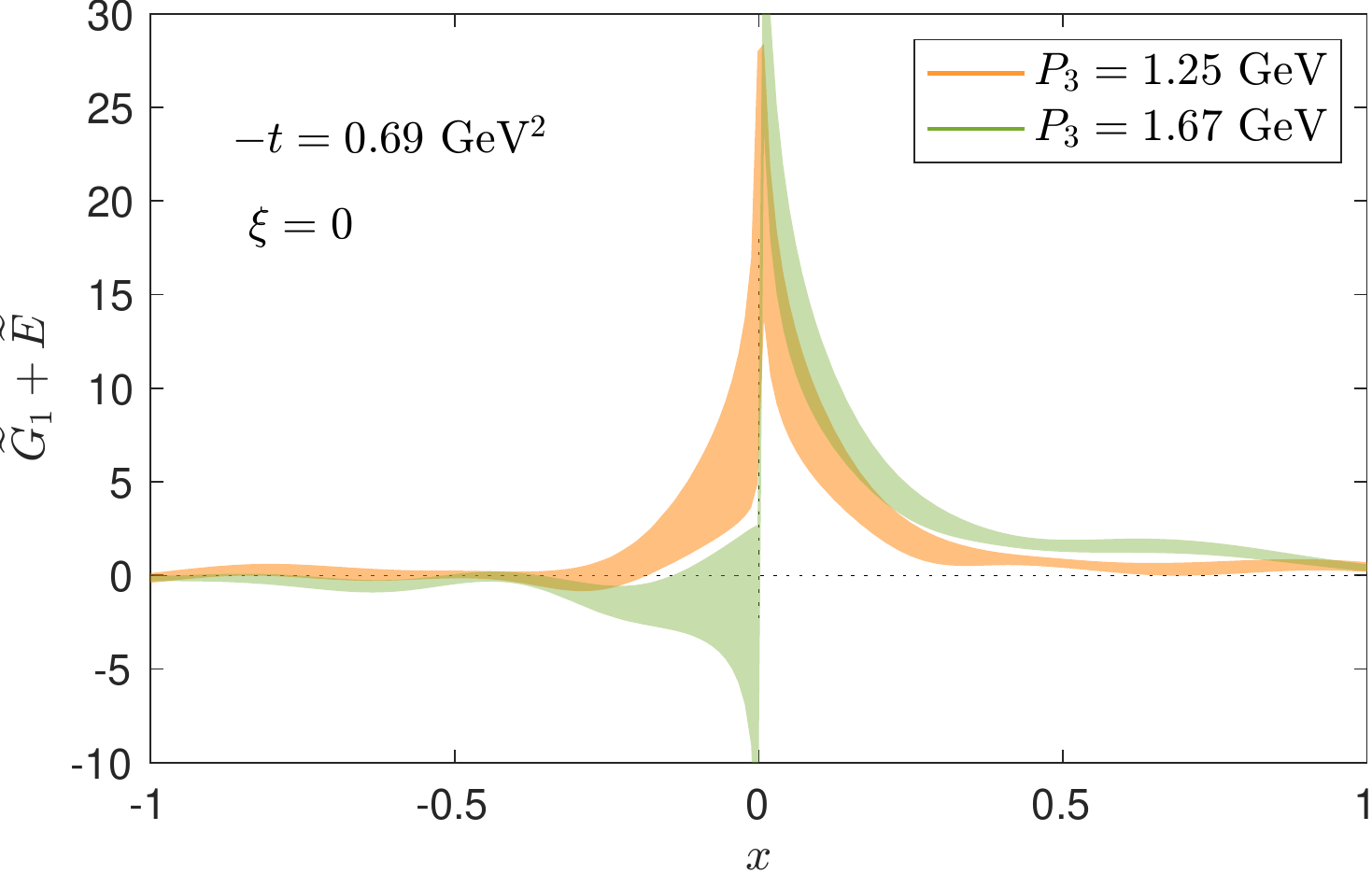}
\caption{Momentum dependence of $\widetilde{H}+\widetilde{G}_2$ (left) and $\widetilde{E}+\widetilde{G}_1$ (right) for $\lbrace \xi=0,-t=0.69$ GeV$^2\rbrace$. The orange and green bands correspond to $P_3~=~1.25$ GeV and $P_3~=~1.67$ GeV, respectively, and indicate only statistical uncertainties.}
\label{fig:GPDs_mom_dep}
\end{figure}

For the axial case, we reconstruct the $x$-dependence from the decomposed matrix elements and apply the matching kernel to obtain the light-cone GPDs as a function of $x$. We remind the reader that we use the results of Ref.~\cite{Bhattacharya:2020xlt}, which correspond to the $g_T$ PDF. Since we only obtain the GPDs at zero skewness, it is anticipated that the matching formalism is the same as for PDFs~\cite{Liu:2019urm}. Here, we focus on the ${\widetilde{H}} + {\widetilde{G}_2}$ and ${\widetilde{E}} + {\widetilde{G}_1}$ GPDs, for which a signal is found. Their $P_3$-dependence is shown in Fig.~\ref{fig:GPDs_mom_dep} for the two highest momenta. We find that the GPDs are very close to each other, with a marginal agreement in the small-$x$ region. Note that the bands correspond to statistical uncertainties only. An investigation of various systematic effects is required before reaching any conclusions. We emphasize that, presently, lattice QCD calculations are not reliable in extracting the small-$x$ region ($x\le 0.15$), nor the antiquark region.

\begin{figure}[h!]
\hspace*{0.2cm}
\includegraphics[scale=0.52]{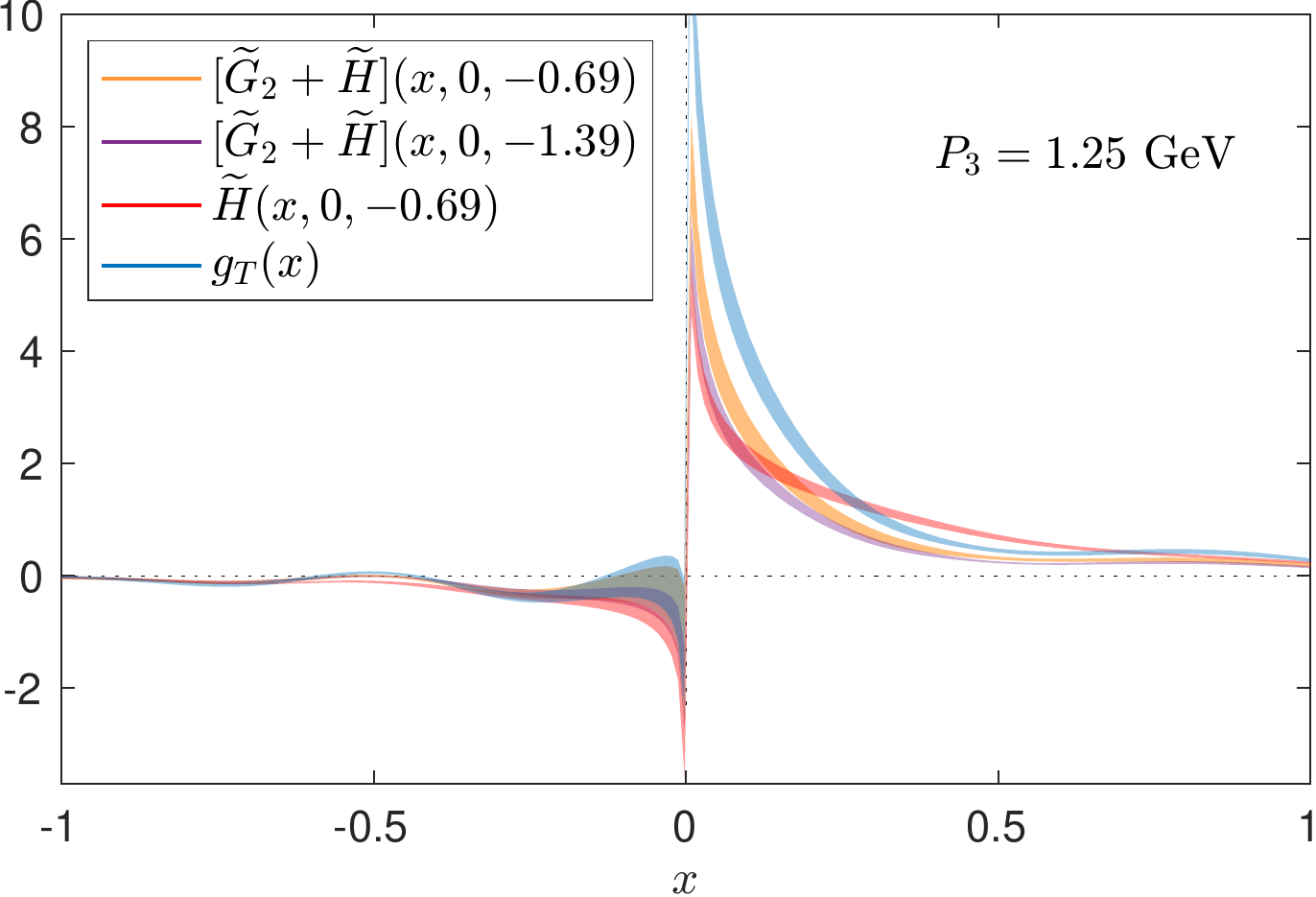} \hspace{0.1cm}
\includegraphics[scale=0.52]{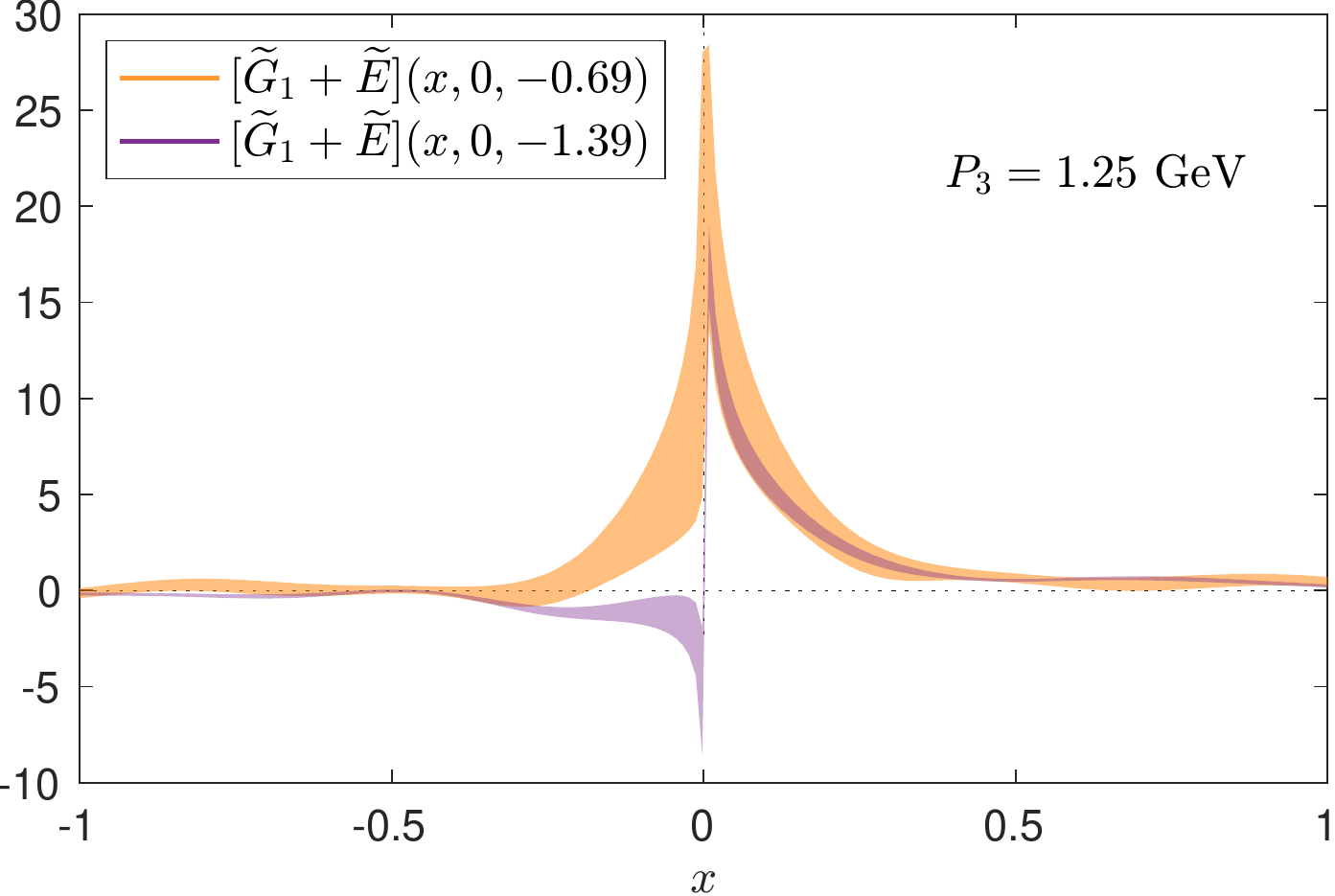}
\caption{Left: $\widetilde{H}+\widetilde{G}_2$ at $\lbrace \xi=0,-t=0.69$ GeV$^2\rbrace$ (orange), $\lbrace \xi=0,-t=1.39$ GeV$^2\rbrace$ (purple) together with the twist-2 $\widetilde{H}$ at $\lbrace \xi=0,-t=0.69$~GeV$^2\rbrace$ (red), and $g_T$ (blue). Right: $\widetilde{E}+\widetilde{G}_1$ at $\lbrace \xi=0,-t=0.69$~GeV$^2\rbrace$ (orange) and $\lbrace \xi=0,-t=1.39$ GeV~$^2\rbrace$ (purple).  The momentum boost for all data is $P_3~=~1.25$~GeV. The bands correspond to statistical uncertainties.}
\label{fig:GPDs_final}
\end{figure}

Finally, in Fig.~\ref{fig:GPDs_final} we compare the twist-3 GPDs at two different values of $t$ with their twist-2 counterparts calculated in Ref.~\cite{Alexandrou:2020zbe}. For the case of $\widetilde{H}+\widetilde{G}_2$ we also compare with their forward limit, $g_T$, which we calculated in a separate work~\cite{Bhattacharya:2020cen}. The setup corresponds to $P_3=1.25$ GeV and $\xi=0$. We can note that $g_T(x)$ is the dominant distribution in magnitude, while ${\widetilde{H}} + {\widetilde{G}_2}$ is similar in magnitude to ${\widetilde{H}}$ at $t=-0.69$ GeV$^2$. Our preliminary results show a mild dependence on $t$ for both ${\widetilde{H}} + {\widetilde{G}_2}$ and ${\widetilde{E}} + {\widetilde{G}_1}$. 

\section{Future work}
We have presented preliminary results on the axial twist-3 GPDs $\widetilde{H}+\widetilde{G}_2,\,\widetilde{E}+\widetilde{G}_1,\,\widetilde{G}_3,\,\widetilde{G}_4$, for three values of $P_3$ and two values of the momentum transfer, that is $-t=0.69,\,1.39$ GeV$^2$. We also showed very preliminary results for the vector matrix elements, that can potentially lead to the extraction of ${H}+{G}_2,\,{E}+{G}_1,\,{G}_3,\,{G}_4$. We find that the signal is reasonable, and for the axial case, we can disentangle the four GPDs. We note that the noise-to-signal ratio is increased for the matrix elements of twist-3 compared to the twist-2 counterparts.

In the near future, we will increase statistics for the current values of $P_3$. We will also include more values of $t$, once the matching formalism is available for $\xi\ne 0$. Another direction is to study ensembles with larger volume, so we can have a more dense range of $t$. Further, we will continue our study for the vector twist-3 GPDs, and also analyze data for the chiral-odd twist-3 case. 

\vspace*{1cm}
\centerline{\textbf{Acknowledgements}}
The work of S.B.~and A.M.~has been supported by the National Science Foundation under grant number PHY-2110472.  A.M.~has also been supported by the U.S. Department of Energy, Office of Science, Office of Nuclear Physics, within the framework of the TMD Topical Collaboration. K.C.\ is supported by the National Science Centre (Poland) grant SONATA BIS no.\ 2016/22/E/ST2/00013. M.C., J.D. and A.S. acknowledge financial support by the U.S. Department of Energy, Office of Nuclear Physics, Early Career Award under Grant No.\ DE-SC0020405. F.S.\ was funded by by the NSFC and the Deutsche Forschungsgemeinschaft (DFG, German Research
Foundation) through the funds provided to the Sino-German Collaborative Research Center TRR110 “Symmetries and the Emergence of Structure in QCD” (NSFC Grant No. 12070131001, DFG Project-ID 196253076 - TRR 110). Computations for this work were carried out in part on facilities of the USQCD Collaboration, which are funded by the Office of Science of the U.S. Department of Energy. 
This research was supported in part by PLGrid Infrastructure (Prometheus supercomputer at AGH Cyfronet in Cracow).
Computations were also partially performed at the Poznan Supercomputing and Networking Center (Eagle supercomputer), the Interdisciplinary Centre for Mathematical and Computational Modelling of the Warsaw University (Okeanos supercomputer), and at the Academic Computer Centre in Gda\'nsk (Tryton supercomputer). The gauge configurations have been generated by the Extended Twisted Mass Collaboration on the KNL (A2) Partition of Marconi at CINECA, through the Prace project Pra13\_3304 ``SIMPHYS".
Inversions were performed using the DD-$\alpha$AMG solver~\cite{Frommer:2013fsa} with twisted mass
  support~\cite{Alexandrou:2016izb}.

\bibliographystyle{JHEP}
{\small
\bibliography{references}}  

\end{document}